\documentclass[12pt]{article}
\usepackage{amssymb}
\usepackage{graphicx}
\usepackage{amsmath}
\newcommand{\be}{\begin{equation}}
\newcommand{\ee}{\end{equation}}
\begin{document}
\baselineskip18pt
\title{L\'{e}vy flights,  dynamical  duality and fractional quantum mechanics}
\author{Piotr Garbaczewski\thanks{Presented at the 21 Marian Smoluchowski Symposium on Statistical Physics}\\
Institute of Physics,  University  of Opole, 45-052 Opole, Poland}
\maketitle
\begin{abstract}

  We discuss   \it  dual \rm  time evolution scenarios which,
  albeit   running according to the same  \it real time \rm   clock,  in each considered case
  may be mapped among  each other by means of a suitable    analytic continuation in time procedure.
    This  dynamical duality   is  a  generic  feature of  diffusion-type processes.  Technically that
    involves a familiar  transformation from a non-Hermitian  Fokker-Planck operator to the
     Hermitian operator (e.g. Schr\"{o}dinger Hamiltonian), whose
    negative is known to generate a dynamical semigroup.
Under suitable restrictions upon the generator, the semigroup admits an analytic continuation in time and ultimately
 yields dual motions. We   analyze  an extension of the duality concept to
  L\'{e}vy flights, free  and  with an external forcing, while  presuming that the  corresponding evolution rule
  (fractional dynamical  semigroup)  is a   dual   counterpart  of the  quantum motion (fractional unitary
  dynamics).
  \end{abstract}
     \noindent
 PACS numbers: 02.50.Ey, 05.20.-y, 05.40.Jc
\vskip0.2cm

\section{Brownian motion inspirations}

\subsection{Diffusion-type processes  and dynamical semigroups }

The Langevin equation for  a one-dimensional stochastic diffusion
process in an  external conservative  force field $F = - (\nabla V)$:
$ \dot{x} = F(x) + \sqrt{2D} b(t)$, where $b(t)$ stands for the normalized white noise $\langle b(t)\rangle =0$,
 $\langle b(t')b(t)\rangle=  \delta (t-t')$,  gives  rise to the corresponding Fokker-Planck equation
 for the probability density  $\overline{\rho }(x,t)$:
 \begin{equation}
 \partial _t\overline{\rho }= D \Delta \overline{\rho } - \nabla ( F \overline{\rho }) \, .
 \end{equation}
 By  means of a standard substitution  $\overline{\rho }(x,t) = \Psi (x,t) \exp[- V(x)/2D]$, \cite{risken},  we  pass to
a generalized diffusion equation for an auxiliary function $\Psi (x,t)$:
 \begin{equation}
 \partial _t \Psi = D \Delta \Psi - {\cal{V}} (x) \Psi
 \end{equation}
 where a  compatibility condition  $ {\cal{V}} (x) = (1/2)[(F^2/2D) +  \nabla  F]$ needs to be respected.  This transformation
 assigns the role of the dynamics  generator to   the  Hermitian (eventually self-adjoint) operator $-\hat{H} = D\Delta - V$.

 Under suitable restrictions upon $V(x)$, $-\hat{H}$ becomes  a legitimate generator of a contractive
  dynamical semigroup $\exp(-\hat{H}t)$, $t\geq 0$. If additionally the dynamical semigroup is amenable to an analytic continuation in time,
   the  contractive  semigroup operator  $\exp(-\hat{H}t)$   can be related  with the unitary operator $\exp(-i\hat{H}t)$ via so-called
   Wick rotation $t\rightarrow it$.
  This  duality observation underlies our forthcoming discussion and generalizations to  L\'{e}vy flights framework.

\subsection{Free propagation and its analytic continuation in time}

The standard  theory of  Gaussian diffusion-type processes takes the Wiener process as the
"free noise" model, with the  Laplacian as  the "noise" generator. It is an element of  folk lore  that the related dissipative
semigroup dynamics  $\exp (tD\Delta )= \exp (-t\hat{H}_0)$  (and thus the heat  equation)   can be mapped into the unitary
dynamics $\exp(itD\Delta ) = \exp( -it\hat{H}_0)$ (and thus the free   Schr\"{o}dinger equation), by means of an analytic continuation
in time procedure, \cite{gar}.
A parameter $D$ may be interpreted dimensionally as $D=\hbar /2m$,  or $D=k_BT/m\beta$ (Einstein's
fluctuation-dissipation statement).

Quite often, this mapping is represented by a formal $it\rightarrow t $ time transformation of the free
 Schr\"{o}dinger picture dynamics  (one should be aware that to
execute a mapping for concrete solutions, a  proper  adjustment of
the time interval boundaries is  necessary):
\begin{equation}
i\partial _t\psi = -D\triangle \psi  \longrightarrow \partial _t
\theta _* =D\triangle \theta _*  \, ,
\end{equation}
 where the notation $\theta _*$
for solutions of the heat equation  has been  adopted,   to stay in
conformity with the forthcoming more general discussion,  where $\theta _*(x,t)$ needs not to be a probability density,
\cite{gar}-\cite{zambrini1}.

The mapping is usually exemplified in terms of integral  kernels $g$ and $k$ as  follows, c.f.
also \cite{wang}:
\begin{equation}
\psi (x,t) = \int dx'g(x-x',t)
\psi (x',0)
\end{equation}
$$
g(x-x',t) \doteq k(x-x',it) = (4\pi iDt)^{-1/2} \, exp[ -\, {(x-x')^2\over
 4iDt}]
$$
and
\begin{equation}
 \theta _*(x,t)= \int dx'k(x-x',t)\theta _*(x',0)
\end{equation}
$$k(x-x',t)\doteq g(x-x',-it) =(4\pi Dt)^{1/2}\, exp[-\, {(x-x')^2\over 4Dt}]\, ,
$$
where the initial $t=0$ data need to be properly  adjusted.
Here, $g(x-x',t)$ is an integral kernel of the unitary evolution operator:
$[\exp(iDt\Delta )\, \psi ](x,0)= \psi (x,t)$.
The heat kernel $k(x-x',y)$ plays the same role with respect to the contractive semigroup operator:
$[\exp(Dt\Delta )\, \theta _*](x,0)$.

The special choice of
\begin{equation}
\psi (x,0) =   (\pi \alpha ^2)^{-1/4} \exp \left( - {\frac{x^2}{2\alpha ^2}}\right)
\end{equation}
implies
\begin{equation}
\psi (x,t) = \left( {\frac{\alpha ^2}{\pi }} \right)^{1/4} (\alpha ^2 + 2iDt)^{-1/2} \exp\left[
 - {\frac{x^2}{2(\alpha ^2+2iDt)}}\right]
\end{equation}
and
\begin{equation}
 \theta _*(x,t)  \doteq  \psi (x,-it) =\left( {\frac{\alpha ^2}{\pi }} \right)^{1/4} (\alpha ^2 + 2Dt)^{-1/2} \exp\left[
 - {\frac{x^2}{2(\alpha ^2+2Dt)}}\right]\, '
\end{equation}
with  $ \theta _*(x,0)= \psi (x,0)$.

 We note that $\rho = |\psi |^2 = \psi \psi
^*$  is a  quantum mechanical  probability density on $R$ for all times
\begin{equation}
\rho (x,t)  =
 \left[ {\frac{\alpha ^2}{\pi (\alpha ^4 + 4D^2t^2)}}\right]^{1/2}
\exp\left[ -{\frac{\alpha ^2x^2}{\alpha ^4 + 4D^2t^2}}\right]  \, .
\end{equation}

The real  solution  $\theta _*(x,t)$ of the heat equation  is not a probability density $\overline{\rho }(x,t)= \theta _*(x,t) \, \, \theta (x,t) $,
unless multiplied   by an appropriate  real function  $\theta (x,t)$ which solves the time adjoint heat equation (that becomes an  ill-posed
dynamical  problem if  considered carelessly).

{\it Case 1:} Since  $\overline{\rho }(x,t) =
[2\pi (\alpha ^2 + 2Dt)]^{-1/2} \exp [-x^2/2(\alpha ^2 +2Dt)]$ actually   is  an example of  the  free   Brownian motion probability density
for all  $t\geq 0$,   we  infer
 \begin{equation}
 \overline{\rho }(x,t) =  (4\pi \alpha ^2)^{1/4}\,   \theta _*(x,t) \doteq  (\theta \,  \theta _*)(x,t)
 \end{equation}
where  $\theta (x,t) \equiv  \theta =(4\pi \alpha ^2)^{1/4} $ is interpreted as a trivial (constant) solution of the time adjoint heat
 equation $\partial _t \theta = - D \Delta \theta $.  We stress that $\theta _* = (4\pi \alpha ^2)^{-1/4} \overline{\rho } \sim \overline{\rho } $.
 This, looking  redundant observation, will prove quite useful  in  a more general
  framework  to be introduced in below.

 {\it Case 2:}  A complex conjugate  $\psi ^*(x,t)=  \psi (x, -t)$  of $\psi (x,t)$, Eq. (7),   solves the time-adjoint Schr\"{o}dinger equation
 $i\partial _t \psi ^* = D\Delta \psi ^*$. Hence  a time-symmetric approach to the analytic continuation
 in time  might look  more compelling.
 Indeed
 \begin{equation}
 \theta (x,t)\doteq \psi ^*(x,it) = \left( {\frac{\alpha ^2}{\pi }} \right)^{1/4} (\alpha ^2 - 2Dt)^{-1/2} \exp\left[
 - {\frac{x^2}{2(\alpha ^2 -2Dt)}}\right]
\end{equation}
is a legitimate solution of the time-adjoint heat equation $\partial _t \theta = - D\Delta \theta $ as long as
$t\in [-T/2,+T/2]$ where $T=\alpha ^2/D$.

In the present case, both time adjoint equations set well defined Cauchy problems (at least in the just defined time interval). The subtle point is that
 the would be  "initial" data for the backward in time evolution, in fact  need to be   the terminal data, given at  the end-point $T/2$ of the
 considered time-interval.

 The only propagation tool, we have in hands, is the heat kernel  (3): $k(x-x',t\rightarrow t-t')$ with $t\geq t'$. There holds
 $\theta _*(x,t) = \int k(x-x',t-t')\, \theta _*(x',t')\,  dx'$ and $\theta (x',t')= \int \theta (x,t)\, k(x-x',t-t')\, dx$
  for any $t'<t \in [-T/2,+T/2]$.

The original quantum mechanical probability density $\rho = |\psi |^2 = \psi \psi
^*$, Eq. (7),  is mapped  into  a  Brownian
bridge (pinned  Brownian motion) probability density:
\begin{equation}
\rho (x,\pm it)  \doteq  \overline{\rho } (x,t) = (\theta \theta _*)(x,t)  =
 \left[ {\frac{\alpha ^2}{\pi (\alpha ^4 - 4D^2t^2)}}\right]^{1/2}
\exp\left[ -{\frac{\alpha ^2x^2}{\alpha ^4 - 4D^2t^2}}\right]  \, .
\end{equation}
The price paid for the time-symmetric appearance  of this formula  is a limitation of the admissible  time span to a finite
 time-interval of length $T=\alpha ^2/D$.

{\it Case 3:} To make a direct  comparison of  {\it Case 2}  with the previous {\it Case 1}, let us confine the time interval
of {\it Case 2} to $[0, +T/2]$.
Now, a  conditional Brownian motion connects $ \overline{\rho }(x,0) = \rho (x,0) =  (\alpha ^2\pi )^{-1/2} \exp (- x^2/\alpha ^2)$ with
$\overline{\rho }(x,t\rightarrow +T/2)$ of Eq. (10). Because  of $T=  \alpha ^2/D $, as $t\rightarrow T/2$, instead of a regular function we
arrive at the linear functional (generalized function), here represented by the Dirac delta  $\delta (x)$. Note that $\delta (x-x')$ is a
 standard initial  $t=0$ value of the heat kernel $k(x-x',t)$.

 This behavior is faithfully reproduced by the time evolution of $\theta _*(x,t)$ and $\theta (x,t)$ that compose $\overline{\rho }(x,t) =
 (\theta _*\, \theta )(x,t)$ for $t\in [0,T/2]$. The initial value of $\theta _*(x,0)=\psi (x,0)$, Eq. (6), is propagated {\it forward}
 in accordance with Eq. (8) to $\theta _*(x,T/2)= (4 \pi \alpha ^2)^{-1/4} \exp(-x^2/4\alpha ^2)$.

In parallel,  $\theta (x,t)$ of (11) interpolates {\it backwards} between  $\theta (x,T/2) \equiv (4\pi \alpha ^2)^{1/4} \, \delta (x)$
and  $\theta (x,0)=\theta _*(x,0)$.  We have here employed  an identity $\delta (a x) = (1/|a|) \delta (x)$. Because of   $f(x)\delta (x) \equiv
f(0) \delta (x)$, we arrive at  $\overline{\rho }(x,T/2) = (\theta _*\, \theta )(x,T/2) \equiv \delta (x)$.

\subsection{Schr\"{o}dinger's  boundary data problem}

The above discussion  provides particular solutions to  so-called Schr\"{o}dinger boundary data problem, under an assumption that
a  Markov  stochastic  process which  interpolates between two a priori  given probability densities $\rho (x,0)$  and $\rho (x,T/2)$ can
be modeled by means of the Gauss probability law (e.g. in terms of the heat kernel).
 That incorporates the free Brownian motion (Wiener process)  and all its conditional variants, Brownian bridges being included,
 \cite{zambrini, zambrini1} and   \cite{blanch}-\cite{klauder}, c.f. also \cite{gar}.

For our purposes
the relevant information is that, if the interpolating process is to
display the Markov property,  then it has to be specified by the joint probability measure
($A$ and $B$ are Borel sets in $R$):
\begin{equation}
m(A,B)= \int_A dx\int_B dy \, m(x,y)
\end{equation}
where ${\int_R m(x,y) dy = \rho (x,0),}$ and   $\int_R m(x,y) dx = \rho (y,T/2)$.
From the start,  we assign   densities to all measures to be dealt with, and  we assume  the
 functional form of the density $m(x,y)$
\begin{equation}
{m(x,y) = f(x)k(x,0,y,T/2)g(y)}
\end{equation}
to  involve  two unknown  functions $f(x)$ and $g(y)$ which are of the same
sign and nonzero, while $k(x,s,y,t)$ is any bounded strictly positive
(dynamical semigroup)  kernel defined for all times  $0\leq s<t\leq T/2$.
For each concrete  choice of the kernel, the  above  integral equations are  known to determine  functions $f(x),\, g(y)$ uniquely
(up to constant factors).

By denoting $\theta _*(x,t)=\int f(z) k(z,0,x,t) dz $ and
$\theta (x,t)=
\int k(x,t,z,T/2)g(z) dz$ it follows that
\begin{equation}
{\overline {\rho}(x,t) = \theta (x,t)\theta _*(x,t) = \int p(y,s,x,t)
\overline {\rho }(y,s) dy ,}
\end{equation}
$$p(y,s,x,t) = k(y,s,x,t){{\theta (x,t)}\over \theta (y,s)},$$
for all $0\leq s<t\leq T/2$. The above $p(y,s,x,t)$ is the transition probability density of the pertinent Markov process that interpolates
between $\overline {\rho}(x,0)$  and $\overline {\rho}(x,T/2)$.
 Cases 1 through 3  are particular examples of  the above reasoning, once $k(x,s,y,t)$  is specified  to be the heat kernel (3) and the corresponding
  boundary density data  are chosen. Clearly, $\theta ^*(x,0)= f(x)$ while $\theta (x,T/2)=g(x)$.

  We recall that  in case of the free evolution,  by  setting $\theta (x,t)= \theta \equiv const $, as in Case 1,  we effectively  transform
   an integral kernel $k$  of the $L^1(R)$   norm-preserving semigroup  into a transition probability density $p$ of the
    Markov stochastic process. Then $\theta ^* \sim \overline{\rho }$.

\section{Free noise models:  L\'{e}vy flights  and   fractional  (L\'{e}vy) semigroups}

The  Schr\"{o}dinger  boundary data problem is amenable to an immediate  generalization to infinitely divisible probability laws which
 induce   contractive semigroups  (and their  kernels)  for   general  Gaussian and  non-Gaussian noise models. They  allow  for   various jump and
  jump-type  stochastic  processes instead of  a  diffusion process.

   A subclass  of stable probability laws  contains a
  subset that is   associated in the literature  with the concept of   L\'{e}vy flights.
  At this point  let us  invoke a functional analytic lore, where contractive semigroup operators, their generators and  the
   pertinent integral  kernels can be directly deduced from  the L\'{e}vy-Khitchine formula, compare e.g. \cite{klauder}.

Let us consider semigroup generators (Hamiltonians, up to dimensional constants) of the form
$\hat{H}=F(\hat{p})$, where  $\hat{p}=-i \nabla $ stands for the momentum  operator  (up to the disregarded $\hbar $ or $2mD$ factor)
  and  for $-\infty <k<+\infty $,  the function  $F=F(k)$ is   real valued,
bounded from below and  locally integrable.
 Then,
\begin{equation}
\exp(-t\hat{H})=\int_{-\infty }^ {+\infty } \exp[-tF(k)] dE(k)
\end{equation}
where $t\geq 0$  and  $dE(k)$ is the spectral measure of $\hat{p}$.

Because of
\begin{equation}
(E(k)f)(x)=
{1\over {\sqrt {2\pi }}}\int_{-\infty }^{k} \exp(ipx) \tilde{f}(p) dp
\end{equation}
where  $\tilde{f}$  is the Fourier transform of $f$, we learn that
\begin{equation}
[\exp(-t\hat{H})]f(x) = [\int_{-\infty }^{+\infty } \exp(-tF(k)) dE(k)f](x)=
\end{equation}
$${{1\over {\sqrt {2\pi }}}\int_{-\infty }^{+\infty }\exp[-tF(k)]
{d\over {dk}}
[\int_{-\infty }^{k} \exp(ipx) \tilde{f}(p) dp]dk = }  \eqno (22)$$
$${1\over {\sqrt {2\pi }}}\int_{-\infty }^{+\infty } \exp(-tF(k)) \exp(ikx)
\tilde{f}(k) dk = [\exp(-tF(p)) \tilde{f}(p)]^{\vee }(x)$$
where the superscript $\vee $ denotes the inverse Fourier transform.

Let us set
\begin{equation}
k_t={1\over {\sqrt {2\pi }}}[\exp(-tF(p)]^{\vee }\, .
\end{equation}
Then the
action of $\exp(-t\hat{H})$ can be given in terms of a convolution (i.e. by means of an integral kernel $k_t\equiv  k(x-y,t)=k(y,0,x,t) $):
\begin{equation}
\exp(-t\hat{H})f = [\\exp(-tF(p)) \tilde{f} (p)]^{\vee } = f*k_t
\end{equation}
where
\begin{equation}
(f*g)(x): =\int_R g(x-z)f(z)dz \, .
\end{equation}

 We shall restrict considerations only to those $F(p)$ which give rise to
 positivity preserving semigroups:  if $F(p)$ satisfies the celebrated
 L\'{e}vy-Khintchine formula, then $k_t$ is a positive measure for all
 $t\geq 0$.
The most general case refers to a  combined  contribution from three types of
processes:  deterministic, Gaussian, and  the jump-type process.

We recall that a characteristic function of a random variable $X$  completely determines a probability distribution of that variable.
If this distribution admits a density we can write $E[\exp(ipX)] =  \int_R \overline{\rho}(x) \exp(ipx) dx$ which,
  for infinitely divisible probability laws,  gives rise to  the
 L\'{e}vy-Khintchine formula
\begin{equation}
E[\exp(ipX)] =   \exp \{ i\alpha p - (\sigma ^2/2)p^2  +  \int_{-\infty }^{+\infty } [exp(ipy) - 1 -
{\frac{ipy}{1+y^2}}] \nu (dy)\}
\end{equation}
where $\nu (dy)$ stands for the so-called L\'{e}vy measure. In terms of Markov stochastic processes all that amounts to a decomposition of $X_t$
into
\begin{equation}
X_t = \alpha t  + \sigma B_t + J_t + M_t
\end{equation}
where $B_t$ stands for the free Brownian motion (Wiener process), $J_t$ is a Poisson process while $M_t$ is a general jump-type process
(more technically, martingale with jumps).

By disregarding the deterministic and jump-type   contributions in the above, we are left with the Wiener process  $X_t= \sigma B_t$.
For a  Gaussian   $\overline{\rho }(x)= (2\pi \sigma ^2)^{-1/2} \exp (-x^2/2\sigma ^2)$ we directly evaluate
$E[\exp(ipx)]= \exp(-\sigma ^2 p^2/2)$.

Let us set $\sigma ^2 = 2Dt$. We get $E[\exp(ipX_t)]= \exp(-tDp^2)$ and subsequently, by employing $p \rightarrow \hat{p}=-i \nabla $,  we
arrive at the contractive semigroup operator $\exp(tD\Delta )$ where the one-dimensional Laplacian $\Delta = d^2/dx^2$ has been introduced.
That amounts to choosing  a special version of the previously introduced Hamiltonian $\hat{H}= F(\hat{p})= D\hat{p}^2$. Note that we can get read
of the constant $D$ by rescaling the time parameter in the above.

Presently, we shall concentrate on the integral part of the L\'{e}vy-Khintchine formula,
which is responsible for arbitrary stochastic jump features. By disregarding the deterministic and Brownian motion entries we arrive at:
\begin{equation}
{F(p) = -   \int_{-\infty }^{+\infty } [exp(ipy) - 1 -
{ipy\over {1+y^2}}]
\nu (dy)}
\end{equation}
where $\nu (dy)$ stands for the appropriate  L\'{e}vy measure. The corresponding non-Gaussian Markov process is characterized by
\begin{equation}
E[\exp(ipX_t)]= \exp[-t F(p)]
\end{equation}
with $F(p)$, (22).  Accordingly, the contractive  semigroup generator  may be defined as follows: $F(\hat{p})= \hat{H}$.

For concreteness we can mention some explicit examples of non-Gaussian  Markov semigroup  generators.
  $F(p) = \gamma   |p|^{\mu }$ where  $\mu <2$ and $\gamma >0$ stands for the intensity parameter
 of the L\'{e}vy  process, upon $p \rightarrow  \hat{p}= -i\nabla $   gives  rise to a pseudo-differential operator
 $\hat{H}=  \gamma  \Delta ^{\mu /2}$ often named the  fractional Hamiltonian. Note that, by construction, it is a positive operator (quite alike
  $-D \Delta $).

The corresponding jump-type   dynamics is interpreted  in terms of  L\'{e}vy flights. In particular
\begin{equation}
F(p)= \gamma  |p| \rightarrow \hat{H}= F(\hat{p}) =  \gamma |\nabla | \doteq \gamma (|\Delta | )^{1/2}
\end{equation}
refers to the Cauchy process.

Since we know that the probability density of the free Brownian motion is  a solution of the Fokker-Planck (here, simply - heat) equation
\begin{equation}
\partial _t\overline{\rho } =  D \Delta \overline{\rho }
\end{equation}
it is instructive to set in comparison the pseudo-differential Fokker-Planck equation which  corresponds to the fractional Hamiltonian and the
fractional semigroup $\exp(-t\hat{H})=\exp(-\gamma |\Delta |^{\mu /2})$
\begin{equation}
\partial _t\overline{\rho } = -  \gamma |\Delta |^{\mu /2} \overline{\rho } \, .
\end{equation}
As mentioned in the discussion of   Case 1, instead of $\overline{\rho }$ in the above we can  insert   $\theta _* \sim \overline{\rho }$,
while remembering that $\theta \equiv const$.

\section{Free fractional Schr\"{o}dinger equation}

Fractional Hamiltonians  $\hat{H}=  \gamma |\Delta |^{\mu /2}$  with  $\mu <2$ and $\gamma >0$  are self-adjoint operators  in
suitable  $L^2(R)$ domains. They are also  positive operators, so that  the respective fractional  semigroups    are holomorphic,
 i. e.   we can replace the time parameter
$t$ by a complex one $\sigma =t+is, \,t>0$  so that
\begin{equation}
 \exp(-\sigma \hat{H})=
\int_R \exp(-\sigma F(k))\, dE(k) \, .
\end{equation}
  Its  action is defined by
\begin{equation}
{[\exp(-\sigma \hat{H})]f = [(\tilde{f}\exp(-\sigma F)]^{\vee } = f*k_{\sigma }}\, .
\end{equation}

Here, the integral  kernel reads $k_{\sigma }={1
\over {\sqrt {2\pi }}}[\exp(-\sigma F)]^{\vee }$. Since $\hat{H}$ is
selfadjoint, the
limit $t\downarrow 0$ leaves us with the unitary group $\exp(-is\hat{H})$,
acting in
the same way: $[exp(-is\hat{H})]f = [\tilde{f} exp(-isF)]^{\vee }$, except
that now
$k_{is}: = {1\over {\sqrt {2\pi}}}[exp(-isF)]^{\vee }$ in general is
\it not
\rm a  probability  measure.

 In view of  unitarity, the unit ball in $L^2$ is an
invariant of the dynamics. Hence  probability  densities, in a standard form $\rho = \psi ^*\, \psi $  can be associated with
solutions of the  free   fractional (pseudodiferential) Schr\"{o}dinger  equations:
\begin{equation}
 i\partial _t \psi (x,t) =   \gamma |\Delta |^{\mu /2} \psi (x,t)
\end{equation}
with  initial data  $\psi (x,0)$. Attempts towards formulating the fractional quantum mechanics can be found in Refs.
 \cite{klauder,laskin,laskin1,cufaro}.

All that amounts to an analytic continuation in time, in close affinity with  the Gaussian pattern (1):
\begin{equation}
i\partial _t \psi  =   \gamma  |\Delta |^{\mu /2} \psi  \longleftrightarrow
\partial _t\theta ^* = -  \gamma |\Delta |^{\mu /2} \theta ^*
\end{equation}
We  assume  that $\theta ^* \sim \overline{\rho}$ and thence the corresponding $\theta \equiv const$.

Stable stochastic processes and their quantum counterparts are plagued by a common disease: it is extremely hard,
if possible at all, to produce insightful analytic solutions.  To get a flavor of intricacies to be faced and the level of technical
difficulties, we shall reproduce some observations in regard  to the Cauchy dynamical  semigroup and its unitary (quantum) partner.
For convenience we scale out a parameter $\gamma $.

For the Cauchy process, whose generator is $|\nabla |$, we deal with a
 probabilistic classics:
\begin{equation}
{\overline {\rho }(x,t) = {1\over {\pi }}\, {t\over {t^2 + x^2}}
\Longrightarrow  k(y,s,x,t) = {1\over {\pi }}{{t-s}\over {(t-s)^2 +
(x-y)^2}}]}
\end{equation}
where $0<s<t$.  We have
$\langle \exp[ipX(t)]\rangle := \int_R \exp(ipx) \overline {\rho }(x,t)\,
dx = \exp[-tF(p)] = \exp(-|p|t)$  and
\begin{equation}
 \overline {\rho }(x,t)=
\int _R   k(y,s,x,t)\,\overline {\rho }(y,s) \, dy
\end{equation}
 for all $t>s\geq 0$.  We recall that $lim_{t\downarrow 0} {\frac{t}{\pi (x^2 +t^2)}}\equiv \delta (x)$.

The characteristic function of $k(y,s,x,t)$ for $y,s$ fixed,  reads
$\exp[ipy - |p|(t-s)]$, and the L\'{e}vy measure needed to evaluate the
L\'{e}vy-Khintchine integral reads:
\begin{equation}
{\nu _0(dy): = lim_{t\downarrow 0} [{1\over t}k(0,0,y,t)]dy ={{dy}
\over
{\pi y^2}}}\, .
\end{equation}

To pass to a dual Cauchy-Schr\"{o}dinger dynamics, we need to perform an analytic continuation in time.   We deal with a holomorphic
fractional semigroup  $\exp(-\sigma t|\nabla |)$, $\sigma = t+ is$, (27). It is clear that  $exp(-t|\nabla |)$ and $exp(-is|\nabla |)$
have a common, identity operator limit as $t\downarrow 0$ and $s\equiv t\downarrow 0$.

An analytic continuation of the Cauchy kernel by means of (28) gives rise to:
\begin{equation}
{k_t(x)={1\over {\pi }} {t\over {x^2+t^2}}\, \longrightarrow  } \,
g_s(x)\doteq  k_{is}(x)={1\over 2}[\delta (x-s) + \delta (x+s)] + {1\over {\pi }}
{\cal P}{is
\over {x^2-s^2}} \, ,
\end{equation}
where ${\cal{P}}$ indicates that a convolution of the integral  kernel with any function should be considered as  a principal value
of an improper integral, \cite{klauder}.
This should be compared with an almost trivial outcome of the previous mapping (2)$\rightarrow $ (3).
Here, we employ  the usual notation for the Dirac delta functionals,
and the
new time label $s$ is a remnant of the limiting procedure $t\downarrow 0 $
in $\sigma =t+is$.

 The function  denoted by $is/\pi (x^2-s^2)$ comes
from the
inverse Fourier transform of $-{i\over \sqrt {2\pi }}sgn (p) sin (sp)$.
Because of
\begin{equation}
{[sgn (p)]^{\vee } = i\sqrt {2\over \pi } {\cal P}({1\over x})}
\end{equation}
where ${\cal P}({1\over x})$ stands for the functional defined in
terms of a
principal value of the integral. Using the notation $\delta _{\pm s}$
for the Dirac delta functional $\delta (x \mp s)$:
\begin{equation}
{[sin (sp)]^{\vee } = i\sqrt {\pi \over 2} (\delta _s - \delta _{-s})}
\end{equation}
we realize that
\begin{equation}
{{1\over \pi }{is\over {x^2-s^2}} = {i\over {2\pi }}(\delta _s -
\delta _{-s})*{\cal P}({1\over x})}
\end{equation}
is given in terms of the implicit  convolution of two generalized functions.
Obviously,  a propagation of an initial function $\psi _0(x)$ to time $t>0$:
\begin{equation}
\psi (x,t)= \int _R g(x-x',t)\psi_0(x') dx'
\end{equation}
 gives a solution of the fractional (Cauchy) Schr\"{o}dinger equation $i\partial _t\psi = - |\nabla | \psi $.

In comparison with the Gaussian case of Section 1, one important difference must be emphasized. The improper integrals, which appear
while evaluating  various convolutions, need to be handled by means of their principal value.
Therefore, a simple $it \rightarrow t $  transformation recipe
no longer  works on the level of integral kernels and respective $\psi $ and $\theta ^*$ functions.

One explicit example is provided by the incongruence of (31) and (34) with respect to  the formal $t\rightarrow -it$ mapping.
Another is provided by considering specific solutions of pseudo-differential equations (30).

To that end, let us consider
$\theta _{*0}(x)= (2/\pi )^{1/2} {\frac{1}{1+x^2}}$, together with $\theta = (2\pi )^{-1/2}$. Then,
$\theta \, \theta _*(x,0) = {\frac{1}{\pi (1+x^2)}}$ is an $L(R)$ normalized Cauchy density, while  $\theta _{*0}(x)$
 itself  is  the  $L^2(R)$ normalized function.  Clearly:
\begin{equation}
\theta _*(x,t) = [\exp(-t|\nabla |)\theta _{*0}](x)=    \int k(y,0,x,t) \theta _*(y,0) dy =
\left( {\frac{2}{\pi }}\right)^{1/2} {\frac{1+t}{x^2 + (1+t)^2}}
\end{equation}
while the corresponding $\psi (x,t)$ with $\psi _0(x) = \theta _{*0}(x)$ reads (for details see e.g. \cite{klauder}):
\begin{equation}
\psi (x,s) = [\exp(-is|\nabla |)\, \psi _0](x) = {1\over 2} [\psi _0(x+s) +\psi _0(x-s)] +
\end{equation}
$$
{i\over 2} [(x-s)\psi_0(x-s) -(x+s)\psi _0(x+s)]\, .
$$

\section{Dynamical duality in external potentials: fractional Schr\"{o}dinger semigroups and L\'{e}vy flights}

\subsection{Schr\"{o}dinger semigroups for Smoluchowski processes}

Considerations of Section 1, where the free quantum dynamics and  free  Brownian motion were considered as dual dynamical scenarios,
can be generalized to an externally perturbed dynamics, \cite{gar}.
Namely, one knows that the  Schr\"{o}dinger equation  for a quantum particle in an external potential $V(x)$,   and the generalized
 heat equation   are connected by analytic continuation in time, known to
take the Feynman-Kac (holomorphic semigroup) kernel into the Green
function of the corresponding quantum mechanical problem.
\begin{equation}
i\partial _t\psi = -D\Delta \psi   +   {\cal{V}} \psi  \longleftrightarrow \partial _t \theta _* = D\Delta \theta _*  -
{\cal{V}}\theta _* \, .
\end{equation}
 Here ${\cal{V}} \doteq V(x)/2mD$.

For $V=V(x), x\in R$, bounded from below, the generator $\hat{H}=-2mD^2
\triangle + V$ is essentially selfadjoint on a natural dense subset of
$L^2$, and the kernel $k(x,s,y,t)=[\exp[-(t-s)\hat{H}]](x,y)$ of the related
dynamical semigroup $\exp(-t\hat{H})$  is strictly positive. The quantum unitary dynamics
$\exp(-i\hat{H}t)$ is  the  an obvious  result of the analytic continuation in time of a dynamical semigroup.

  Assumptions concerning the
admissible potential may be relaxed. The necessary demands are that $\hat{H}$ is self-adjoint and bounded from below. Then the respective
 dynamical semigroup is holomorphic.

The key role of an integral kernel of the dynamical semigroup operator  has been elucidated in formulas (11)-(13), where  an explicit form
of a transition probability density of the Markov diffusion process was given. We have determined as well the time development of $\theta _*(x,t)$
and $\theta (x,t)$, so that $\overline{\rho }(x,t)= (\theta \theta _*)(x,t)$ is a probability density of the pertinent process.

If  we a priori consider   $\theta (x,t) $   in the functional form  $\theta (x,t)\doteq \exp \Phi (x,t)$, so that
 $\theta _*(x,t)  \doteq \overline{\rho }(x,t)  \exp[-\Phi (x)]$,  and  properly   define   the forward drift
$b(x,t) \doteq  2D \nabla \Phi (x,t)$ in the  pertinent  Fokker-Planck equation:
\begin{equation}
\partial _t \overline{\rho } = D\Delta \overline {\rho } - \nabla (b\, \overline{\rho })
\end{equation}
we can recast  a diffusion problem  in terms of a pair of time adjoint generalized heat equations
\begin{equation}
\partial _t\theta _*= D\Delta \theta _* - {\cal{V}}\theta _*
\end{equation}
and
\begin{equation}
\partial _t \theta = -D\Delta \theta + {\cal{V}} \theta \, ,
\end{equation}
i. e. as the Schr\"{o}dinger boundary data problem, where an interpolating stochastic process is  uniquely determined by
a continuous and positive
Feynman-Kac kernel of the Schr\"{o}dinger semigroup  $\exp(-t\hat{H})$, where  $\hat{H} = -D\Delta + {\cal{V}}$.

If our departure point is the Fokker-Planck (or Langevin) equation with the a priori  prescribed   potential function $\Phi (x,t)$
 for the forward drift  $b(x,t)$, then
 the backward equation  (44) becomes  an identity from which ${\cal{V}}$ directly  follows, in terms of $\Phi $ and its derivatives,
  \cite{blanch,olk0}:
\begin{equation}
{\cal{V}}(x,t) = \bigl [\partial _t \Phi \, +\, {1\over 2} ({b^2
\over {2D}}+ \nabla b)\bigr ]
\end{equation}
For the  time-independent drift  potential, which is the case for standard Smoluchowski diffusion processes,  we get (c.f.  also
 \cite{risken}, where  the a transformation  of the  Fokker-Planck equations (42) into an associated Hermitian problem (43) is described in detail):
\begin{equation}
{\cal{V}}(x) = \bigl [{1\over 2} ({b^2
\over {2D}}+ \nabla b)\bigr ] \, .
\end{equation}
Notice that $\Phi (x)$  is defined up to an additive constant.

To give an example of a pedestrian reasoning based on the above procedure in case of a concrete  Smoluchowski diffusion processes,
let us begin from the  Langevin equation for the one-dimensional stochastic
process in the external conservative  force field $F(x) = - (\nabla V)(x)$ (to keep in touch with the previous notations, note that
 $\Phi \equiv -V$):
\begin{equation}
{\frac{dx}{dt}} = F(x) + \sqrt{2D} B(t)
\end{equation}
 where $B(t)$ stands for the normalized white noise: $\langle B(t)\rangle =0$,
 $\langle B(t')B(t)\rangle= \delta (t-t')$.

 The corresponding Fokker-Planck equation
 for the probability density  $\rho (x,t)$ reads:
 \begin{equation}
 \partial _t\overline{\rho }= D \Delta \overline{\rho  }  -\nabla ( F \overline{\rho })
 \end{equation}
 and  by  means of a  substitution  $\overline{\rho }(x,t) = \theta _*(x,t) \exp[- V(x)/2D]$, \cite{risken}, can be
 transformed into the  generalized diffusion equation for an auxiliary function $\theta ^* (x,t)$:
 \begin{equation}
 \partial _t \theta _* = D \Delta \theta _* - {\cal{V}} \theta _*
 \end{equation}
 where the consistency condition (reconciling the functional form of ${\cal{V}}$ with this for $F$)
 \begin{equation}
 {\cal{V}} = {\frac{1}2}\left( {\frac{F^2}{2D }} + \nabla F\right) \, .
 \end{equation}
 directly comes from the time-adjoint equation
 \begin{equation}
\partial _t \theta \equiv 0=-D\Delta \theta + {\cal{V}} \theta \,
 \end{equation}
with $\theta (x) = \exp[-V(x)/2D]$.

For the Ornstein-Uhlenbeck process  $b(x) = F(x) = - \kappa x$ and accordingly
\begin{equation}
  {\cal{V}} (x) = {\frac{\kappa ^2x^2}{4D}} - {\frac{\kappa }2} \, .
\end{equation}
is an explicit  functional form of the  potential ${\cal{V}}$, present in previous  formulas (41)-(44).

\subsection{Fractional semigroups  and perturbed L\'{e}vy flights}

External perturbations in the additive form:
\begin{equation}
 i\partial _t \psi (x,t) =   \gamma |\Delta |^{\mu /2} \psi (x,t)  + {\cal{V}}(x)\psi (x,t)
\end{equation}
were considered in the framework of fractional quantum  mechanics, \cite{laskin}-\cite{cufaro}, c.f. also \cite{klauder,olk}.
With the dual dynamics concept in mind, Eq. (30), we expect that an anlytic continuation in time (if admitted) takes us from the
fractional Schr\"{o}dinger equation  to the  fractional analog of the generalized diffusion equation:
\begin{equation}
\partial _t\theta ^* = -  \gamma |\Delta |^{\mu /2} \theta ^*   -  {\cal{V}} \theta ^*   \, .
\end{equation}
The time-adjoint equation has the form
\begin{equation}
\partial _t\theta = \gamma |\Delta |^{\mu /2} \theta    + {\cal{V}} \theta    \, ,
\end{equation}
We shall be particularly interested in the time-independent $\theta (x,t) \equiv \theta (x)$, an assumption affine to that
involved in the passage from (44)-(46).

Hermitian fractional problems of the  form (48) and/or (49)  have also  been studied in Refs. \cite{brockmann, geisel,geisel1}.
However, the major  (albeit implicit, never openly stated) assumption of Refs. \cite{brockmann, geisel,geisel1}  was to consider
 the  so-called  step L\'{e}vy process instead of the jump-type L\'{e}vy process proper.

  This amounts to introducing a  lower bound on the length of admissible jumps: arbitrarily small jumps are then excluded.
   That allows to by-pass a serious  technical obstacle.  Indeed, for a pseudo-differential operator $\gamma \Delta ^{\mu /2}$,
    the action on a function from  its domain can be greatly simplified by disregarding  jumps of length  \it not \rm  exceeding a
     fixed $\epsilon >0$,   see e.g. Refs. \cite{klauder,olk}:
    \begin{equation}
\gamma |\Delta | ^{\mu /2} f)(x)\, =\, - \int_R [f(x+y) - f(x) - {{y\, \nabla f(x)}
\over {1+y^2}}]\, \nu (dy)
\end{equation}
$$\Downarrow $$
 $$ \gamma | \Delta |_{\epsilon}^{\mu /2} f)(x)\, =\, - \int_{|y|>\epsilon } [f(x+y) - f(x) ] \nu (dy) \, .
$$
Compare e.g. Eq. (2) in \cite{geisel} and Eq. (6) in \cite{geisel1}. Note that these Authors  have skipped  the minus sign that
\it must \rm  appear on the right-hand-side of both  formulas (50).

As a side comment, let us point out that the principal  integral value
issues of Section 3  would not arise in our previous discussion of Cauchy flights and their generators, if arbitrarily small jumps
 were eliminated from the start. Nonetheless, if the $\epsilon \downarrow 0$ limit is under control, the step process
 can be considered as a meaningful approximation of the fully-fledged (perturbed)  jump-type L\'{e}vy process.
 This  approximation
  problem has been investigated  in   detail,   in the construction of the perturbed Cauchy process, governed by the Hermitian dynamical problem
  (53), with the input (55),  under  suitable restrictions on the behavior of $ {\cal{V}}$, \cite{olk}.

Let us come back to time-adjoint   fractional  equations (54) and (55).   We have   $\overline{\rho }(x,t) = (\theta \, \theta ^*)(x,t)$ and
employ the trial ansatz  of Section 4.2:
 \begin{equation}
 \theta (x,t) \equiv \theta (x)= \exp[\Phi (x)]
 \end{equation}
 $$  \theta ^*(x,t) = \overline{\rho }(x,t)\, \exp [- \Phi (x)]\, .
 $$

Accordingly (55)  implies, compare e.g.  \cite{brockmann}  for  an independent argument:
\begin{equation}
{\cal{V}}  =   -\gamma \exp(-\Phi ) |\Delta |^{\mu /2} \exp(\Phi )
\end{equation}
 to be compared with Eq. (8) in Ref. \cite{geisel}. In view of (54) we have
 \begin{equation}
  \partial _t \overline{\rho } = \theta \partial _t \theta ^*= -   \gamma  \exp(\phi ) [|\Delta |^{\mu /2} \exp(-\Phi ) \overline{\rho }]
    +  {\cal{V}} \overline{\rho } \doteq - \nabla j \, .
 \end{equation}

 Langevin-style description  of perturbed L\'{e}vy flights  (deterministic component plus the L\'{e}vy noise contribution) are known,
 \cite{fogedby, ditlevsen,chechkin},  to  generate   fractional Fokker-Planck equations of the form
 \begin{equation}
\partial _t\overline{\rho } =  - \nabla ( F\, \overline{\rho }) - \gamma |\Delta | ^{\mu /2} \overline{\rho } \doteq   - \nabla j
 \end{equation}
where $F=- \nabla V\equiv \nabla \Phi $,  we face   problems which are left  unsettled  at the present stage of our investigation: \\

(i) May  the stochastic processes   driving (58) and/or  (59)   coincide
 under any circumstances,  or basically  not at all ? \\

 (ii)  Give an insightful/useful definition   of the probability current $j(x,t)$ in both
 considered  cases, while remembering that for fractional derivatives the composition rule for consecutive (Riesz) derivatives
 typically breaks down. \\

Both  problems (i) and (ii) have have an immediate resolution in case of diffusion-type processes, where  by departing from
 the Langevin equation one infers  Fokker-Planck  and continuity equations.    In turn,
 these  equations   can be alternatively derived   by  means of the Schr\"{o}dinger boundary
data problem,provided its  integral kernel stems from thee Schr\"{o}dinger semigroup, both in the free and perturbed cases.
The stochastic diffusion process (corresponding to that associated with the  Langevin equation)  is then reconstructed as well.
Thence, the  Schr\"{o}dinger loop  gets closed.

While passing to L\'{e}vy processes, we have demonstrated that, with suitable reservations,  this Schr\"{o}dinger "loop" can be completed
in case of free L\'{e}vy flights. However, the "loop"   remains incomplete (neither definitely proved or disproved) for  perturbed  L\'{e}vy
 flights.

 At this point we should mention clear indications  \cite{brockmann} that, once discussing L\'{e}vy flights, we actually
 encounter two different classes of  processes with incompatible dynamical properties.
 One class is related to the Langevin equation, another - termed topological -
 relies on  the "potential landscape" provided by the effective potential ${\cal{V}}(x)$.
  An extended discussion of the latter problem has been postponed to the forthcoming paper, c.f. \cite{gar1}.

{\bf Acknowledgement:} Partial support from the Laboratory for Physical Foundations of  Information Processing
is gratefully acknowledged.

\end{document}